\documentclass[aps,pre,preprint,groupedaddress]{revtex4-1}

\usepackage{graphicx} 
\usepackage{dcolumn} 
\usepackage{bm}
\newcommand{\ignore}[1]{} 
\usepackage{epsfig} 
\usepackage{color}
\usepackage{amsmath,amssymb}

\begin{document}

\title{Noisy multistate voter model for flocking in finite dimensions}

\author{Ernesto S. Loscar}
\author{Gabriel Baglietto}

\affiliation{Instituto de F\'{i}sica de L\'{i}quidos y Sistemas Biol\'ogicos (IFLYSIB), UNLP, CCT La Plata-CONICET, Calle 59 no.~789, B1900BTE La  Plata, Argentina}

\author{Federico Vazquez}
\email[]{fede.vazmin@gmail.com}

\affiliation{Instituto de C\'{a}lculo, FCEN, Universidad de Buenos Aires and CONICET, Buenos Aires, Argentina}

\date{\today}

\begin{abstract}
We study a model for the collective behavior of self-propelled particles subject to pairwise copying interactions and noise.  Particles move at a constant speed $v$ on a two--dimensional space and, in a single step of the dynamics, each particle adopts the direction of motion of a randomly chosen neighboring particle within a distance $R=1$, with the addition of a perturbation of amplitude $\eta$ (noise).  We investigate how the global level of particles' alignment (order) is affected by their motion and the noise amplitude $\eta$.  In the static case scenario $v=0$ where particles are fixed at the sites of a square lattice and interact with their first neighbors, we find that for any noise $\eta_c>0$ the system reaches a steady state of complete disorder in the thermodynamic limit, while for $\eta=0$ full order is eventually achieved for a system with any number of particles $N$.  Therefore, the model displays a transition at zero noise when particles are static, and thus there are no ordered steady states for a finite noise ($\eta>0$).  We show that the finite-size transition noise vanishes with $N$ as $\eta_c^{\mbox{\tiny 1D}} \sim N^{-1}$ and $\eta_c^{\mbox{\tiny 2D}} \sim \left(N \ln N \right)^{-1/2}$ in one and two--dimensional lattices, respectively, which is linked to known results on the behavior of a type of noisy voter model for catalytic reactions.  When particles are allowed to move in the space at a finite speed $v>0$, an ordered phase emerges, characterized by a fraction of particles moving in a similar direction.  The system exhibits an order-disorder phase transition at a noise amplitude $\eta_c>0$ that is proportional to $v$, and that scales approximately as $\eta_c \sim v \, (-\ln v)^{-1/2}$ for $v \ll 1$.  These results show that the motion of particles is able to sustain a state of global order in a system with voter-like interactions.
\end{abstract}

\maketitle

\section{Introduction}
\label{introduction}

The study of the collective properties of systems composed by self-propelled individuals has been the focus of intense research in the last two decades \cite{vicsek2012, marchetti2013,menzel2015}. The flocking behavior of a large group of animals is observed in many different species such as fish, birds, bacteria and insects, among others. From a statistical physics viewpoint, the interactions between particles in a system are responsible of its collective behavior, and lead to well characterized classes represented by archetype models.  For the case of flocking, the alignment interaction among individuals is usually modeled as a local averaging of moving directions of nearby individuals, plus a noise that accounts for errors in the average process \cite{vicsek-1995}.  A crucial role in the emergent behavior of the system is played by the displacement of the individuals, which changes dramatically its ordering properties \cite{toner-1995}.

Within the context of flocking, the dynamics of collective alignment in groups of fish was recently studied in \cite{jhawar2020}.  The authors performed experiments with cichlid fish \emph{Etroplus suratensis} that swim in a circular shallow tank, in order to explore how schooling is affected by the fish group size.  The level of group alignment is quantified by a vector order parameter ${\bf M}$ that is the average velocity of fish, also called group polarization, in such a way that $|{\bf M}| \sim 1$ corresponds to a polarized state where fish move in a coherent direction, while $|{\bf M}| \sim 0$ represents a collectively disordered state --each fish moving in a random direction.  Performing the experiments for group sizes $N=15$, $30$ and $60$, they found that the collective alignment $|{\bf M}|$ increases as $N$ decreases.  An insight into this phenomenon is given by a phenomenological stochastic differential equation (SDE) for the time evolution of ${\bf M}$, where its parameters were extracted from the experimental data.  It is shown that group polarization is the result of the interplay between the drift and the demographic (population) noise terms in the SDE, that is, the fewer the fish, the greater the demographic noise and so the greater the alignment level.  Thus, they conclude that schooling (highly polarized and coherent motion) is induced by the intrinsic population noise that arises from the stochasticity related to the finite number of interacting fish.  They derived the SDE for ${\bf M}$ by means of a mean-field (MF) model in which particles (fish) interact by pairs and follow a simple imitation dynamics: each particle either copies the direction of another random particle or spontaneously changes its direction, modeled as an external noise of amplitude $\eta$.  They also show that other ternary or higher-order aligning interactions, including local averages like in the Vicsek-like family of models, are unnecessary to explain these experimental results.  Therefore, they arrive to the conclusion that the minimal theoretical mechanism that  reproduces the collective alignment properties of fish observed in the experiments is that of pairwise interactions with copying dynamics and noise.  We notice that the noiseless version of this particular alignment dynamics that induces flocking was first introduced in \cite{baglietto2018}, where the authors study the collective motion of particles on a two--dimensional ($2D$) space subject to voter-like interactions, that is, each particle aligns its direction of motion with that of a random neighboring particle within an interaction radius.  

From the theoretical point of view, an interesting result can be inferred from the work in \cite{jhawar2020} by analyzing the SDE for the group polarization ${\bf M}$.  That is, this equation predicts complete order ($|{\bf M}|=1$) for zero noise ($\eta=0$) and full disorder ($|{\bf M}|=0$) for any finite noise amplitude $\eta>0$ in the $N \to \infty$ limit.  This observation is in agreement with recent analytical results obtained in a similar model with a discrete set of $S$ angular directions, a multistate voter model (MSVM) with external noise \cite{Vazquez-2019}, where it is shown that the order parameter $|{\bf M}|^2$ approaches $1.0$ as $1-|{\bf M}|^2 \sim \eta^2 N$ in the $\eta \to 0$ limit, and vanishes when $N$ increases as $1/(\eta^2 N)$ for any $0 < \eta \ll 1$.  Thus, the partial order obtained with voter interactions and noise in a MF set up is only a finite size effect that eventually disappears in the thermodynamic limit.  These results suggest a peculiar order-disorder transition at zero noise, unseen in related flocking models such as the binary Vicsek model \cite{Chou-2015} where each particle averages its direction with that of other random particle, and the transition happens at a critical noise larger than zero.  However, we notice that the experimental results obtained in \cite{jhawar2020} correspond to fish moving on a $2D$ set up (tank), while both the SDE and the model in \cite{Vazquez-2019} are for a MF set up (infinite dimension), where every particle interacts with any other particle, and thus motion plays no role in the dynamics.  It is natural, therefore, to wonder whether these results hold when particles move on a $2D$ space.  Do space and motion affect the transition at zero noise?

In this article we study a noisy multistate voter model for flocking in finite dimensions, and we investigate the order-disorder phase transition in different case scenarios.  We start by analyzing the simplest case of all-to-all interactions or MF.  We then explore the static case where each particle occupies a site of a square lattice and interacts with its first nearest-neighbors, and we finally study the dynamic case in which particles move on a $2D$ continuous space and change their direction when they interact with other nearby particles that are located within a distance $R=1$.  In  the case that particles are allowed to have only two possible angular states and interact on a MF set up, the model turns to be equivalent to the noisy voter model (NVM) introduced in \cite{Kirman-1993,Granovsky-1995}, in which each individual of a population holds one of two states (opinions) that are updated by either copying the state of a random neighbor or spontaneously switching state (noise).  In the absence of noise, any finite population eventually reaches full order (consensus) in all dimensions, as in the original voter model \cite{clifford1973,holley1975}, with all individuals sharing the same opinion.  However, the addition of a weak noise leads to a bi-stable regime in which the system jumps between two steady states corresponding to a quasi-consensus in one or the other opinion \cite{Kirman-1993,Granovsky-1995}, while for strong noise the system remains disordered.  This is in line with the fact that adding thermal bulk noise in the voter model destroys global order in any dimension \cite{Henkel-2008}, even when the noise is weak.  In square lattices, the NVM is equivalent to a particular limit of the catalytic reaction model with desorption originally introduced in \cite{Fichthorn-1988} and widely studied after \cite{Fichthorn-1989,Considine-1989}, which exhibits a finite size transition induced by noise called {\it saturation}  transition \cite{Clement-1991a,Clement-1991b,Flament-1992}.   More recently, the dynamics of the NVM has been investigated in complex networks \cite{Carro-2016,Peralta-2018-a,Peralta-2018-b}, and its version with multiple states has been explored in fully connected systems \cite{Azque-2019,Vazquez-2019}.  Also, an asymmetric variant of the NVM with long-range interactions has recently been proposed to study the competition between two species for territory \cite{Martinez-2020}.

While in $2D$ lattice models bulk noise inhibits the formation of long range order in the thermodynamic limit, it is known that in flocking systems the displacement of particles plays an ordering role.  This ordering phenomenon is observed in the Vicsek model, thought as a non equilibrium version of the XY model in $2D$ with particles moving ballistically in the directions of their spins.  That is, while the Vicsek model can sustain long-range order for finite values of noise amplitude due to particles' motion 
\cite{toner-1995}, the $2D$ XY model is unable to do so \cite{MWtheorem-1966}.  Then, the velocity of particles in Vicsek-type models leads to steady states associated with a new ordered phase below a transition noise $\eta_c$.  However, voter-type interactions (copying) are different from Vicsek-type interactions (averaging), leading to different behaviors in MF and three dimensions: long-range order in the XY model, and disorder in the NVM.  Therefore, in the flocking voter model (FVM) studied in this article, we expect a non-trivial competition between the ordering mechanism generated by particles' motion and the typical disordering effect induced by noisy voter interactions that leads to complete disorder in the thermodynamic limit.  Thus, we aim to explore whether the ordered phase observed in flocking models is still present in the FVM, or it is rather completely suppressed by noise.

The rest of the article is organized as follows.  In section~\ref{model} we define the model.  Section~\ref{mean-field} presents MF results, while section~\ref{static} is dedicated to the static version of the model in one and two dimensional square lattices.  In section~\ref{dynamic} we study the dynamic version of the model in a continuous $2D$ space.  We investigate the effects of particles' velocity in the transition, with a particular focus on the behavior at low speeds in the thermodynamic limit.  Finally, in section~\ref{conclusions} we summarize and give some conclusions.

\section{The model}
\label{model}

A set of $N$ particles are allowed to move at a constant speed $v$ on a $2D$ square box of side $L$ with periodic boundary conditions.  The position and velocity of particle $i$ ($i=1,2,..,N$) at time $t$ are denoted by $\boldsymbol{r}^{t}_i = (x_i^t,y_i^t)$ and $\boldsymbol{v}^{t}_i=(v \cos \theta_i^t, v \sin \theta_i^t)$, respectively, where $v=|\boldsymbol{v}^{t}_i|$ is the particle's speed and $\theta_i^t$ is its angular moving direction.  The density of particles $\rho = N/L^2$ is fixed at $0.5$ in our analysis, unless stated.  Initially, each particle adopts a random position inside the box and points in a random direction.  In a given time step $\Delta t =1$ of the dynamics, each particle $i$ updates it position and direction according to
\begin{subequations}
  \begin{alignat}{2}
  	\label{r-t}
	\boldsymbol{r} ^{t+1}_i &= \boldsymbol{r}^{t}_i+\boldsymbol{v}^{t}_i \, \Delta t, \\
	\label{theta-t} 
	\theta^{t+1}_i &= \theta^{t}_j+ \xi_i^{t+1},
	\end{alignat}
\end{subequations}
where $\theta^{t}_j$ is the moving direction of a randomly chosen particle $j$ that is inside a disk of radius $R=1$ centered at $\boldsymbol{r}_i^{t}$, and $\xi_i^t$ is a random angle drawn uniformly in $[-\eta \pi,\eta \pi)$ with amplitude $\eta$ ($0<\eta<1$).  This update is performed for all particles at the same time (parallel update).  That is, each particle moves at a constant speed $v$ following a given straight path and updates its direction at integer times $t = 1, 2, 3, ...$, by adopting the direction of a random neighboring particle with an error of amplitude $\eta$.  If a particle has no neighbors inside its interaction range $R$, then its direction is changed only by the noise $\xi$.  

In flocking models, noise --in its various forms-- plays a fundamental role in the behavior of the system.  It is known that the amplitude of noise $\eta$ induces an order-disorder phase transition, from a phase where a large fraction of particles move in a similar direction (order) for small $\eta$, to a phase in which particles move in random directions (disorder) for large $\eta$.  To study this phenomenon in the FVM we define the order parameter (see for instance \cite{jhawar2020,baglietto2018})\begin{equation}
	\varphi(t) \equiv \frac{1}{v \, N} \left| \sum_{i=1}^{N}  \boldsymbol{v}^{t}_i \right| = \frac{1}{N} \sqrt{ \left[\sum_{i=1}^N \cos \theta_i^t \right]^2 +\left[\sum_{i=1}^N \sin \theta_i^t \right]^2}
\label{Eq:phi}
\end{equation}
that measures the level of collective alignment in the system (magnitude of the normalized mean velocity of all particles), and the susceptibility 
\begin{equation}
	\chi \equiv N \left[\langle \varphi^2 \rangle - \langle \varphi
\rangle^2 \right],
\label{Eq:chi}
\end{equation}
which accounts for the amplitude of fluctuations of $\varphi$ at the stationary state.  Here $\langle \varphi^m \rangle$ is the $m$-th moment of $\varphi$, and the symbol $\langle \cdot \rangle$ represents the average value of a given magnitude over many realizations of the dynamics at the steady state.

Our aim is to explore via computational simulations and scaling theory how space and motion affects the phase transition in the FVM.  For that, we first study the model in MF ($R=L$), we then explore the static case $v=0$ in lattices, and we finally investigate the dynamic case $v>0$ in $2D$.

\section{Mean Field}
\label{mean-field}

In order to gain an insight into the behavior of the FVM, we start by analyzing in this section the simplest case scenario of all-to-all interactions or MF, which corresponds to the large interaction range limit $R \to L$ of the model defined in section~\ref{model}.  In this case, the dynamics of the angular directions of particles $\theta$ is independent of the positions of particles, and thus it is entirely determined by Eq.~(\ref{theta-t}).  That is, each particle simply adopts the direction of another randomly chosen particle in the system, with the addition of noise.  This dynamics is equivalent to that of the multistate voter model with imperfect copying introduced and studied in \cite{Vazquez-2019}, in the limit of continuum angular states.  In a single time step $\Delta t =1/N$, a particle $i$ with state $\theta_i$ is picked at random, then it copies the state $\theta_j$ of another randomly chosen particle $j$, and this state is slightly perturbed:
\begin{equation}
	\theta_i (t+\Delta t) =  \theta_j(t) +\xi_i(t+\Delta t).
\label{Eq:dynamicSeq}
\end{equation}
We note that we are implementing here a \emph{sequential update} in which only one particle updates its state in a time step, unlike the \emph{parallel update} where all $N$ particles are updated at once.  However, we have verified that the behavior of the macroscopic variables $\varphi$ and $\chi$ under the parallel update is recovered by making the substitution $N \to 2N$ in the results obtained with the sequential update, as mathematically proved by Blythe and McKane in \cite{Blythe-2007} for population genetic models akin to the voter model.  Inversely, the transformation $N \to N/2$ allows to obtain the behavior under the sequential update from the results with parallel update.
   
\begin{figure}[t]
  \begin{center}
    \vspace{0.5cm}
    \begin{tabular}{cc}
      \hspace{-1.5cm}
      \includegraphics[width=5.3cm, bb=70 -20 550 550]{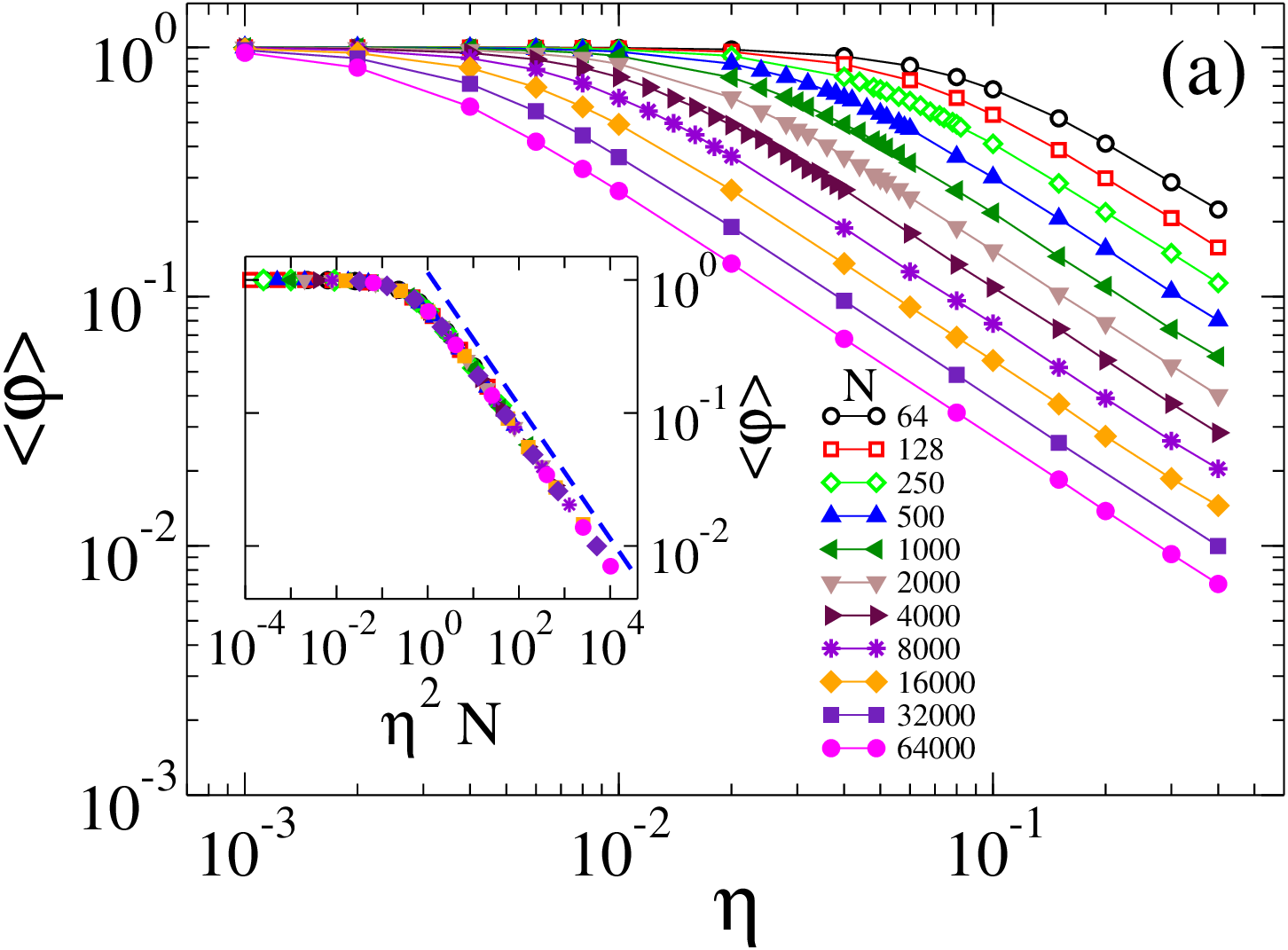} 
      & \hspace{3.0cm}      
      \includegraphics[width=5.3cm, bb=70 -20 550 550]{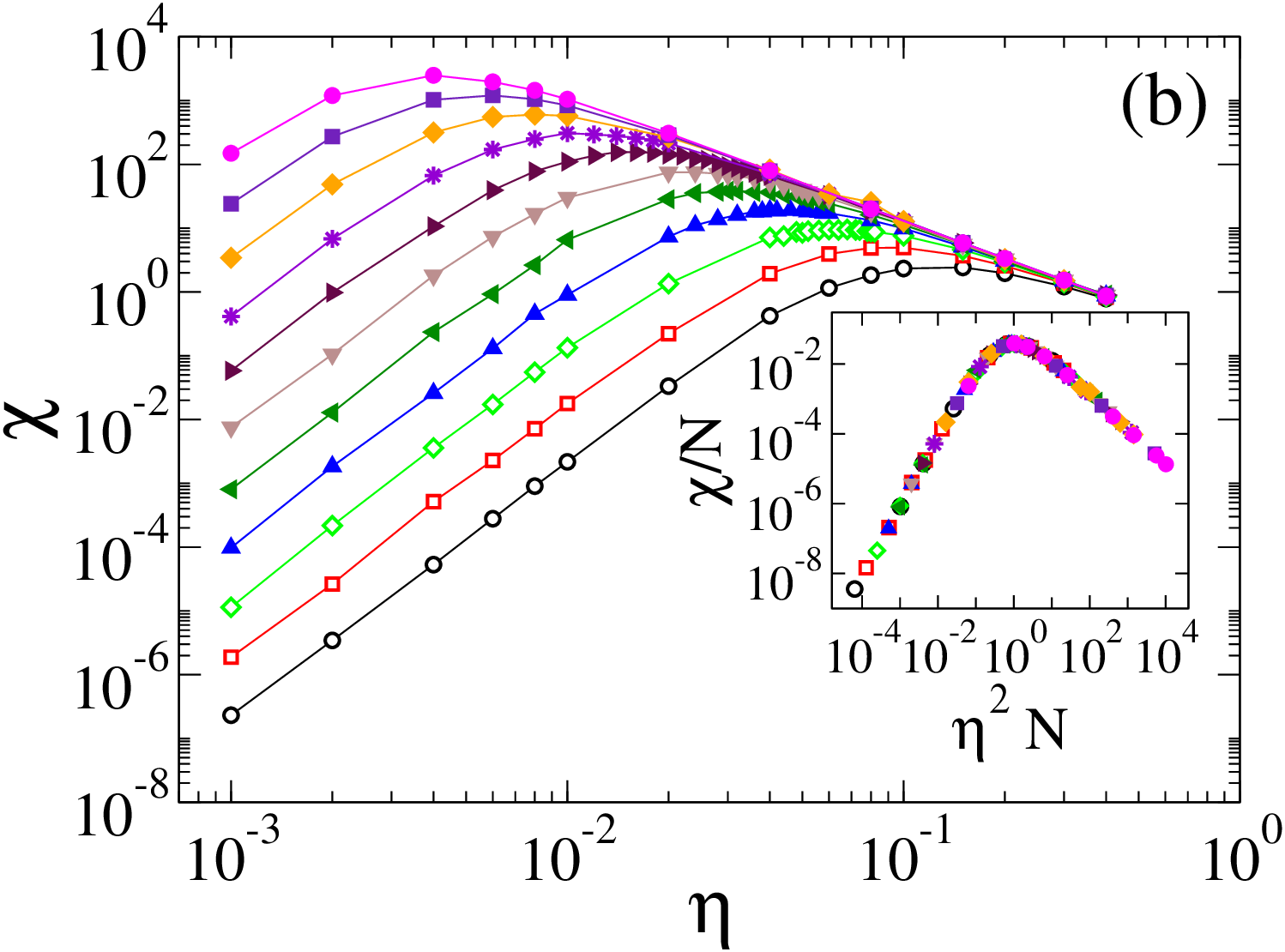} 
    \end{tabular}    
    \caption{Results of the FVM in MF.  (a) Average value of the order parameter $\varphi$ at the stationary state as a function of noise amplitude $\eta$ for the system sizes $N$ indicated in the legend.  The inset shows the collapse of the data points when they are plotted as a function of the scaling variable $x_{\mbox{\tiny MF}} = \eta^2 N$.  The dashed line has slope $-1/2$.  (b) Susceptibility $\chi$ vs $\eta$ for the same system sizes as in panel (a).  Inset: collapse of the data when it is plotted vs $x_{\mbox{\tiny MF}}$ and the $y$--axis is rescaled by $N^{-1}$.  Averages were done in a time window $\Delta t \sim10^7$ over $10$ independent realizations.}
    \label{phi-chi-MF}
  \end{center}
\end{figure}

Figure~\ref{phi-chi-MF} shows simulation results for the model in MF.  Data points correspond to average values in a time interval after the system reached the stationary state, between times $t=10^6$ and $t=2 \times 10^7$, and over $10$ independent realizations.  In panel (a) we observe that the order parameter $\varphi$ continuously decreases as $\eta$ increases, and that approaches the value $\varphi=1$ (full order) as $\eta \to 0$, which corresponds to the absorbing consensus state obtained in the zero noise case $\eta=0$, as it is known from previous works of the multistate voter model 
\cite{Starnini-2012,Pickering-2016,baglietto2018,Vazquez-2019}.  We also see that, for a fixed value of $\eta>0$, $\varphi$ vanishes as the system size $N$ increases, suggesting that $\varphi \to 0$ for any $\eta>0$ in the $N \to \infty$ limit.  Indeed, an expression for the scaling of $\langle \varphi \rangle$ with $\eta$ and $N$ that confirms this assumption can be obtained from analytical results of this model recently presented in \cite{Vazquez-2019}, for an order parameter $\psi=\varphi^2$.  It was shown in \cite{Vazquez-2019} that $\langle \psi \rangle \sim \left(\eta^2 N \right)^{-1}$ for $\eta \ll 1$ and $\eta^2 N \gtrsim 1$, and thus assuming $\langle \varphi \rangle \sim \langle \psi \rangle^{1/2}$ we obtain the approximate MF behavior 
\begin{equation}
	\langle \varphi \rangle_{\mbox{\tiny MF}} \sim \left( \eta^2 N \right)^{-1/2}  ~~~ \mbox{for $\eta \ll1$ and $\eta^2 N \gtrsim 1$}.
	\label{phi-eta-N}
\end{equation}
In the inset of Fig.~\ref{phi-chi-MF}(a) we plot the data as a function of the scaling variable $x_{\mbox{\tiny MF}} \equiv \eta^2 N$, where we can see that $\langle \varphi \rangle_{\mbox{\tiny MF}}$ obeys the power law decay from Eq.~(\ref{phi-eta-N}) for $\eta^2 N \gtrsim 1$ (dashed line).  We also observe a good collapse of the curves for different system sizes in the entire range of $x_{\mbox{\tiny MF}}$, showing that the order parameter is a function of $x_{\mbox{\tiny MF}}$, $\langle \varphi \rangle_{\mbox{\tiny MF}} = f \left( \eta^2 N \right)$, with $f \left( x_{\mbox{\tiny MF}} \right) \sim x_{\mbox{\tiny MF}}^{-1/2}$ for $x_{\mbox{\tiny MF}} \gtrsim 1$.  

\begin{figure}[t]
  \begin{center}
    \vspace{0.5cm}
    \begin{tabular}{cc}
      \hspace{-0.8cm}
      \includegraphics[width=5.7cm, bb=70 -20 550 550]{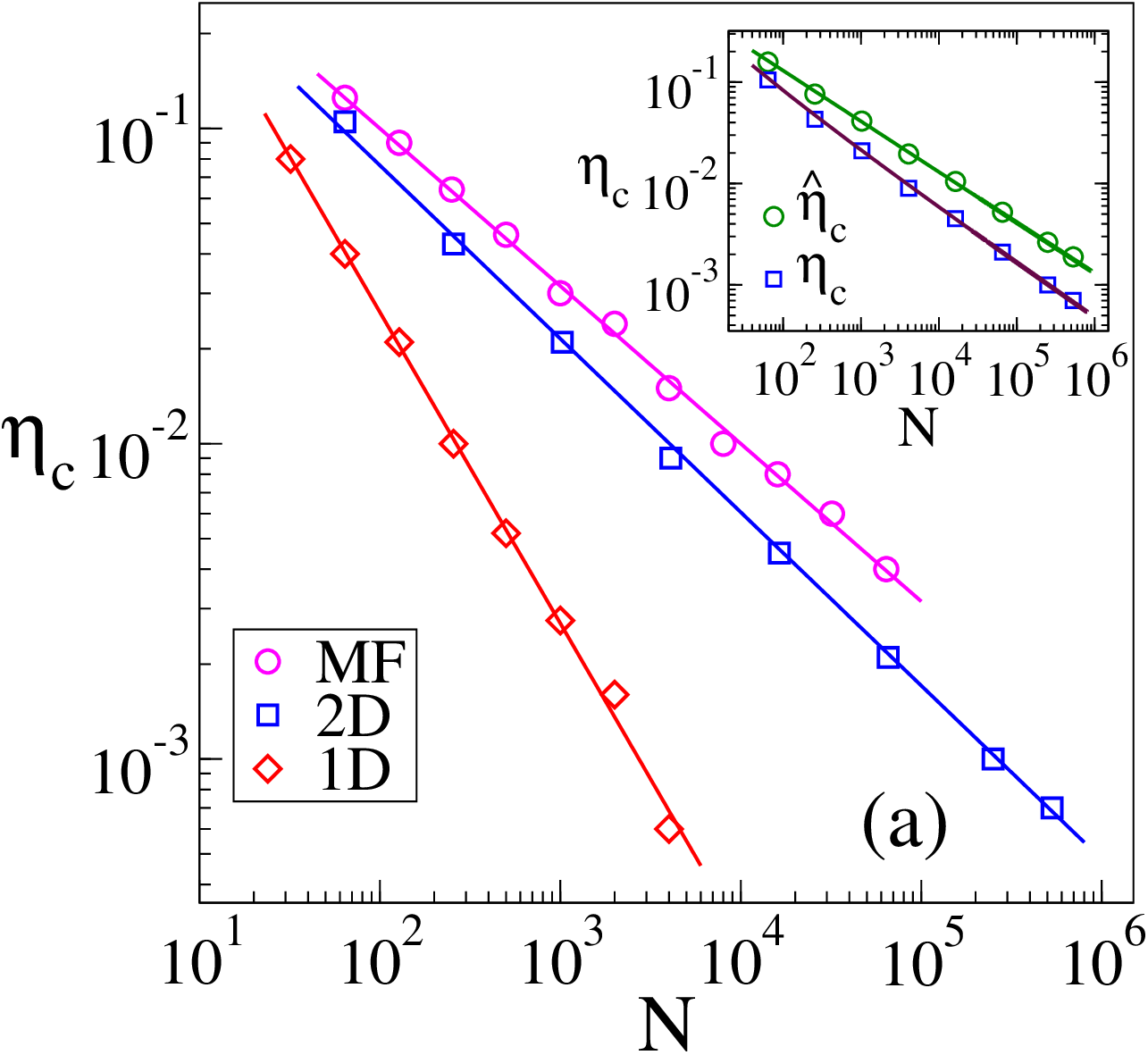} 
      & \hspace{2.5cm}      
      \includegraphics[width=5.7cm, bb=70 -20 550 550]{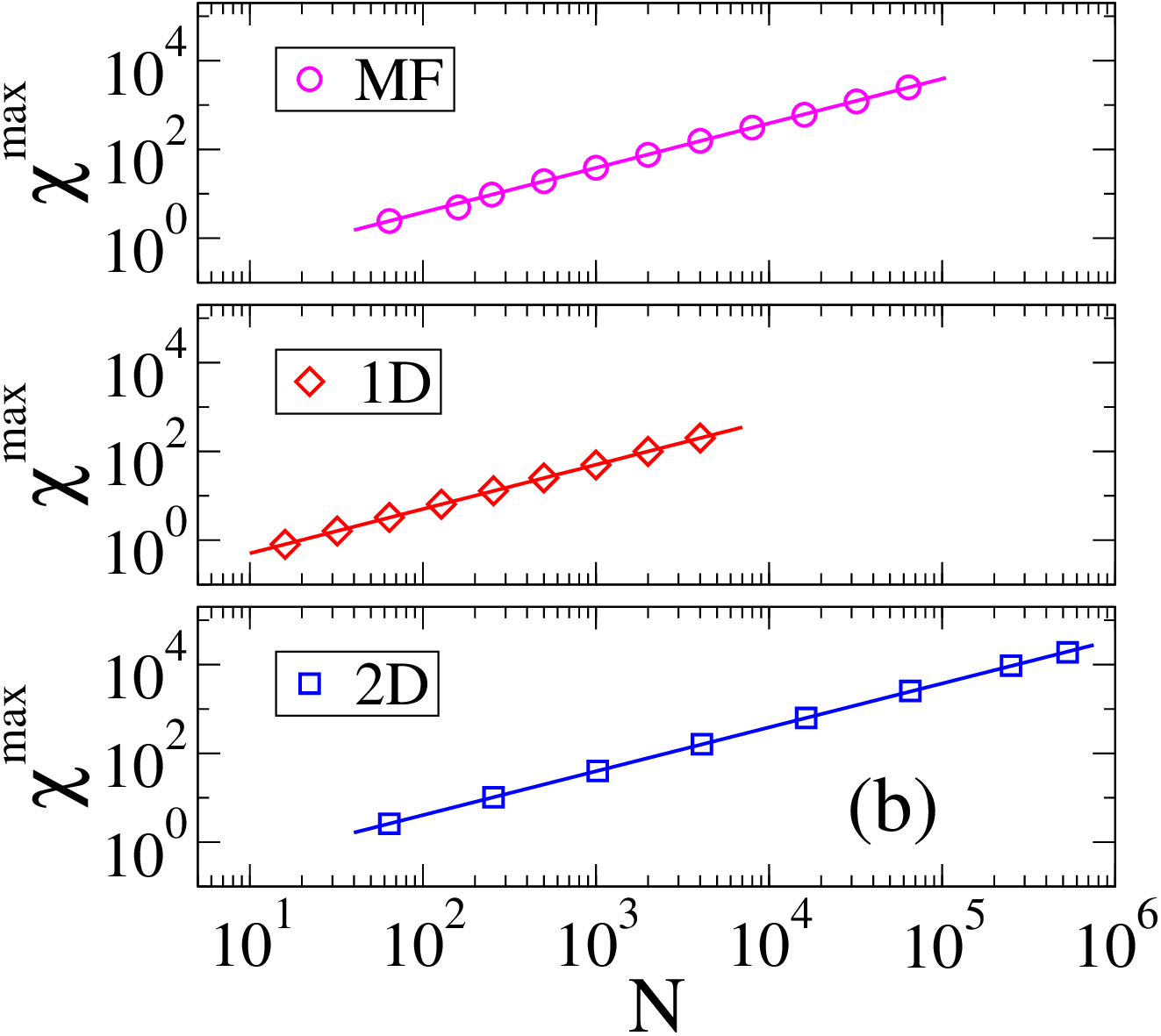} 
    \end{tabular}    
    \caption{Results of the FVM in MF (circles) and in square lattices of dimensions $d=1$ (diamonds) and $d=2$ (squares).  (a). Transition noise $\eta_c$ vs system size $N$.  Straight lines are best power-law fits $\eta_c = A \, N^{-\alpha}$ with exponents $\alpha=0.5 \pm 0.015$ (MF), $0.99 \pm 0.02$ ($d=1$) and $0.56 \pm 0.01$ ($d=2$). Inset: $\eta_c$ for $d=2$ (squares) and the effective noise $\hat{\eta}_c=\eta_c \sqrt{-\ln \eta_c}$ (circles).  The upper solid line is the best power law fit $\hat{\eta}_c \simeq B \, N^{-1/2}$, with $B=1.3 \pm 0.04$, while the bottom solid curve is the approximation $\eta_c \simeq 1.8 \left(N \ln N \right)^{-1/2}$ from Eq.~(\ref{etac-N-lnN}).  (b) Maximum value of the susceptibility $\chi^{\mbox{\tiny max}}$ vs $N$.  Best power-law fits $\chi^{\mbox{\tiny max}} \sim N^{\gamma}$ (straight lines) have exponents $\gamma=1.01 \pm 0.01$ (MF), $0.997 \pm 0.005$ ($d=1$) and $0.99 \pm 0.01$ ($d=2$).}
    \label{etac-chimax}
  \end{center}
\end{figure}

The results above imply that in the absence of noise $\eta=0$ the system reaches full order ($\varphi=1$), but a tiny amount of noise $\eta>0$ is enough to drive the system to complete disorder ($\varphi=0$) in the thermodynamic limit, which suggests a transition at zero noise.  To study this in more detail, we show in Fig.~\ref{phi-chi-MF}(b) the behavior of the susceptibility $\chi$ with $\eta$.  We observe that the curve for a given system size $N$ exhibits a maximum that is an indication of a transition that depends on $N$, between an ordered phase for $\eta<\eta_c(N)$ and a disordered phase for $\eta>\eta_c(N)$, where the transition point $\eta_c(N)$ is estimated as the location of the peak.  In Fig.~\ref{etac-chimax}(a) we plot the transition noise $\eta_c$ vs $N$ (circles), where we can see that $\eta_c$ vanishes as $N$ increases following a power-law behavior $N^{-\alpha}$, with a best fitting exponent $\alpha = 0.5 \pm 0.015$.  This implies a transition value $\eta_c(\infty)=0$ in the thermodynamic limit.  In panel (b) of Fig.~\ref{etac-chimax} we see that the maximum value of the susceptibility increases with $N$ as $\chi^{\mbox{\tiny max}} \sim N^{\gamma}$, where $\gamma \simeq 1.01 \pm 0.01$ is the best fitting exponent.  

These scalings can be nicely verified by assuming that $\chi$ is also a function of the scaling variable $x_{\mbox{\tiny MF}} = \eta^2 N$ for $\varphi$ in Fig.~\ref{phi-chi-MF}.  Indeed, rescaling the $y$--axis of Fig.~\ref{phi-chi-MF}(b) by $N^{-1}$ and plotting the resulting data vs $x_{\mbox{\tiny MF}}$ we find a good collapse of all curves for different $N$ values (see inset), showing that the MF susceptibility behaves as 
\begin{equation}
	\chi_{\mbox{\tiny MF}} = N g \left( \eta^2 N \right),
	\label{chi-eta-N}
\end{equation}
where $g(x_{\mbox{\tiny MF}})$ is a smooth function of $x_{\mbox{\tiny MF}}$.  From Eq.~(\ref{chi-eta-N}) we have that at the MF transition point $\eta_c^{\mbox{\tiny MF}}$ is $\chi^{\mbox{\tiny max}}_{\mbox{\tiny MF}}/N =  g\left[ \left( \eta_c^{\mbox{\tiny MF}} \right)^2 N \right] = \mbox{constant}$ and, therefore, 
\begin{equation}
	\eta_c^{\mbox{\tiny MF}} \sim N^{-1/2},
	\label{etac-N-MF}
\end{equation}	
in agreement with numerical results [Fig.~\ref{etac-chimax}(a)].  

In summary, the mean field version of the FVM exhibits an order-disorder phase transition at zero noise $\eta_c=0$ in the thermodynamic limit, between a perfectly ordered phase where $\varphi=1$ for $\eta=0$ and a completely disordered phase where $\varphi=0$ for 
$\eta > 0$.

\section{Static case $v=0$ in one and two dimensions}
\label{static}

In this section we analyze the static version of the FVM in finite dimensions.  For that, we consider that each particle occupies a site of a square lattice of length $L$ and $d$ dimensions ($N=L^d$ sites), and interacts with its $2^d$ nearest neighbors only.  We have simulated the dynamics of the model under the sequential update described in section~\ref{mean-field} on lattices of dimensions $d=1$ and $d=2$ with periodic boundary conditions.  In a time step $\Delta t=1/N$, a randomly selected particle copies the angular state of a first neighbor chosen at random, with the addition of an error of amplitude $\eta$.

\begin{figure}[t]
  \begin{center}
    \vspace{0.5cm}
    \begin{tabular}{cc}
      \hspace{-1.5cm}
      \includegraphics[width=5.3cm, bb=70 -20 550 550]{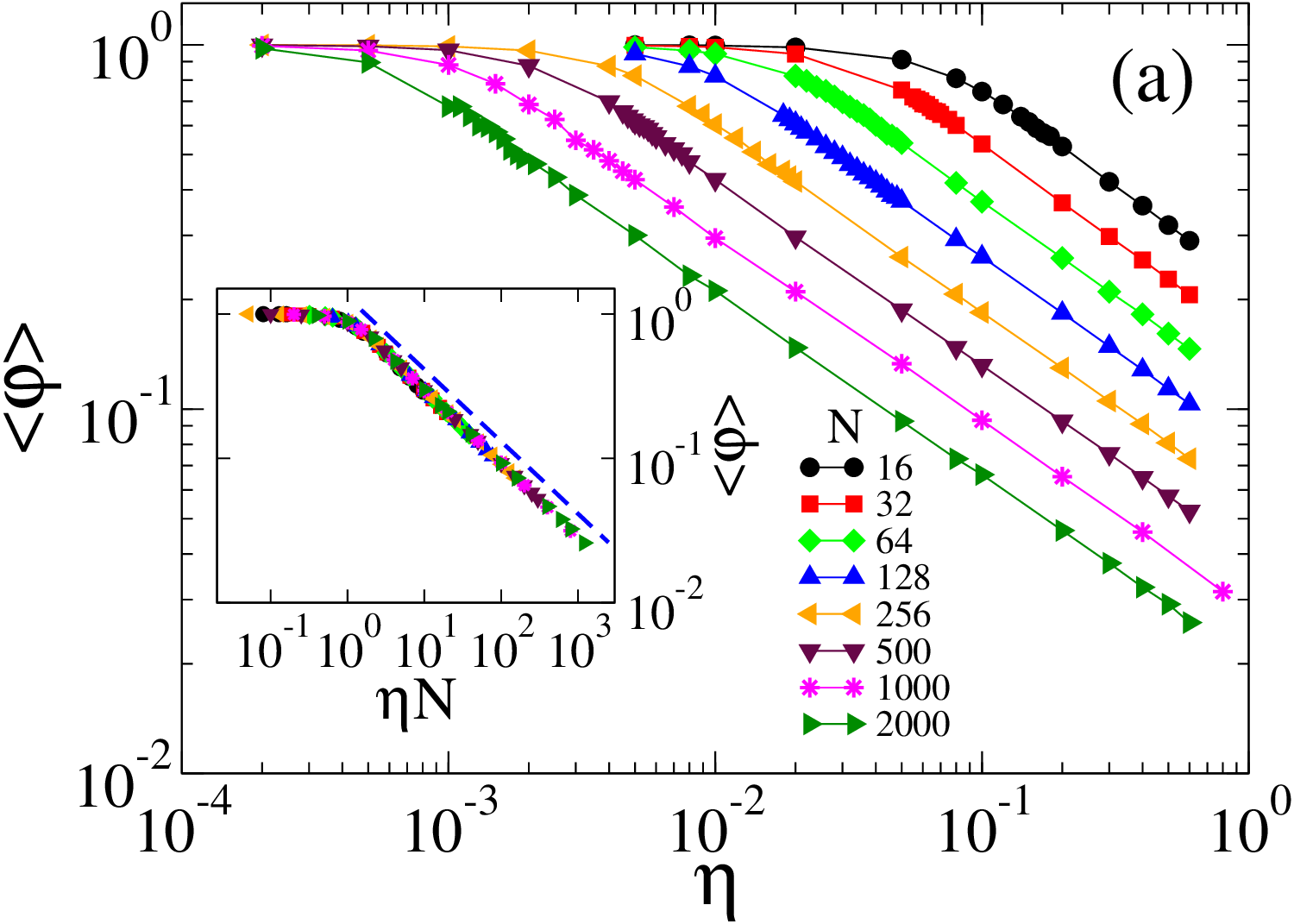} 
      & \hspace{3.0cm}      
      \includegraphics[width=5.3cm, bb=70 -20 550 550]{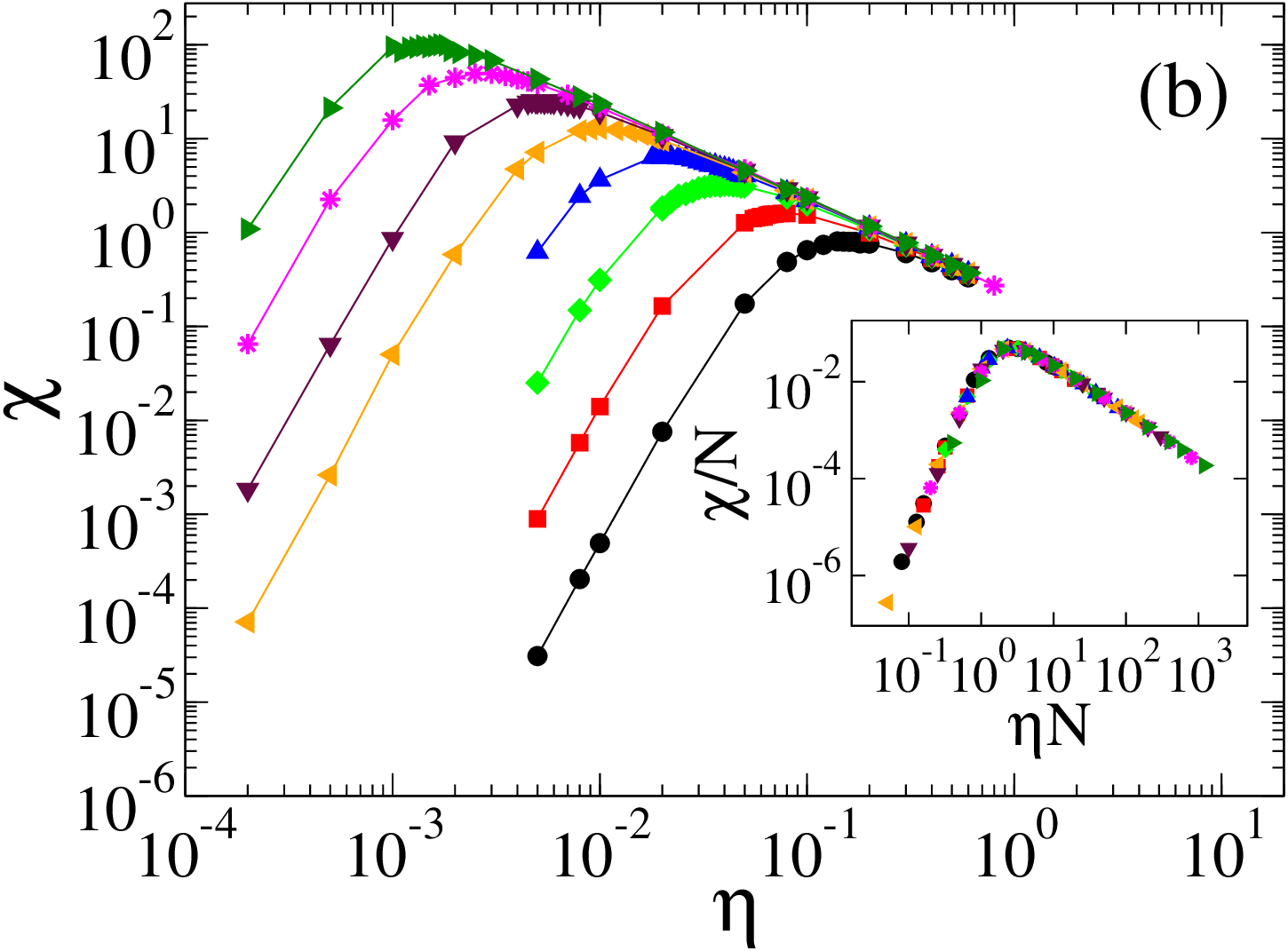} 
    \end{tabular}    
    \caption{Results of the static version of the FVM in one dimension.  (a) Average value of 
    $\varphi$ at the stationary state vs $\eta$ for the system sizes $N$ indicated in the legend.  Inset: same data vs the scaling variable $x_{\mbox{\tiny 1D}} = \eta N$ showing the collapse of curves for different $N$ values.  The dashed line has slope $-1/2$.  (b) Susceptibility $\chi$ vs $\eta$ for the same system sizes as in panel (a).  Inset: $x$ and $y$--axis are rescaled by $N$ and $N^{-1}$, respectively, to show the collapse of the data.}
    \label{phi-chi-1D}
  \end{center}
\end{figure}

Figure~\ref{phi-chi-1D} shows simulation results for the FVM in one dimension.  The behavior of $\langle \varphi \rangle$ and $\chi$ are similar to those of the MF model, with a scaling variable $x_{\mbox{\tiny 1D}} \equiv \eta N$ in this one--dimensional case.  The variable $x_{\mbox{\tiny 1D}}$ was obtained from the behavior of the transition noise $\eta_c^{\mbox{\tiny 1D}}$ with $N$ given by the peak of $\chi$ in panel (b) of Fig.~\ref{phi-chi-1D}.  We found $\eta_c^{\mbox{\tiny 1D}} \sim N^{-\alpha}$, with $\alpha = 0.99 \pm 0.02$ [Fig.~\ref{etac-chimax}(a)], while for the peak of the susceptibility we found the scaling $\chi^{\mbox{\tiny max}} \sim N^{\gamma}$, with $\gamma = 0.997 \pm 0.005$ [Fig.~\ref{etac-chimax}(b)].  Therefore, assuming the scalings 
\begin{eqnarray}
	\label{etac-N-1D}
	\eta_c^{\mbox{\tiny 1D}} &\sim& N^{-1} ~~~ \mbox{and} \\
	\label{chimax-N-1D}
	\chi^{\mbox{\tiny max}} &\sim& N,
\end{eqnarray}	
we arrive at the following scaling for the susceptibility:
\begin{equation}
	\chi_{\mbox{\tiny 1D}} = N g_1( \eta N),
	\label{chi-eta-N-1D}
\end{equation}
and thus the scaling variable is $x_{\mbox{\tiny 1D}}=\eta N$, as stated above.  Indeed, we can check in the insets of Fig.~\ref{phi-chi-1D} the collapse of the curves for different system sizes when the data is plotted vs $x_{\mbox{\tiny 1D}}$, and the $y$--axis in panel (b) is rescaled by $N^{-1}$.  Also, in the inset of panel (a) we show that the order parameter scales as $\langle \varphi \rangle_{\mbox{\tiny 1D}} \sim x_{\mbox{\tiny 1D}}^{-1/2}$ for $x_{\mbox{\tiny 1D}} \gtrsim 1$ (dashed line), which exhibits the same behavior with respect to the scaling variable as that of MF [Eq.~(\ref{phi-eta-N})], i.e., a power-law decay with exponent $1/2$.
 
The scaling relation Eq.~(\ref{etac-N-1D}) shows that the transition noise vanishes with $N$ and, therefore, we conclude that the static version of the FVM in one dimension exhibits an order-disorder transition at zero noise in the thermodynamic limit, as it happens in MF. 

\begin{figure}
	\includegraphics[width=10cm]{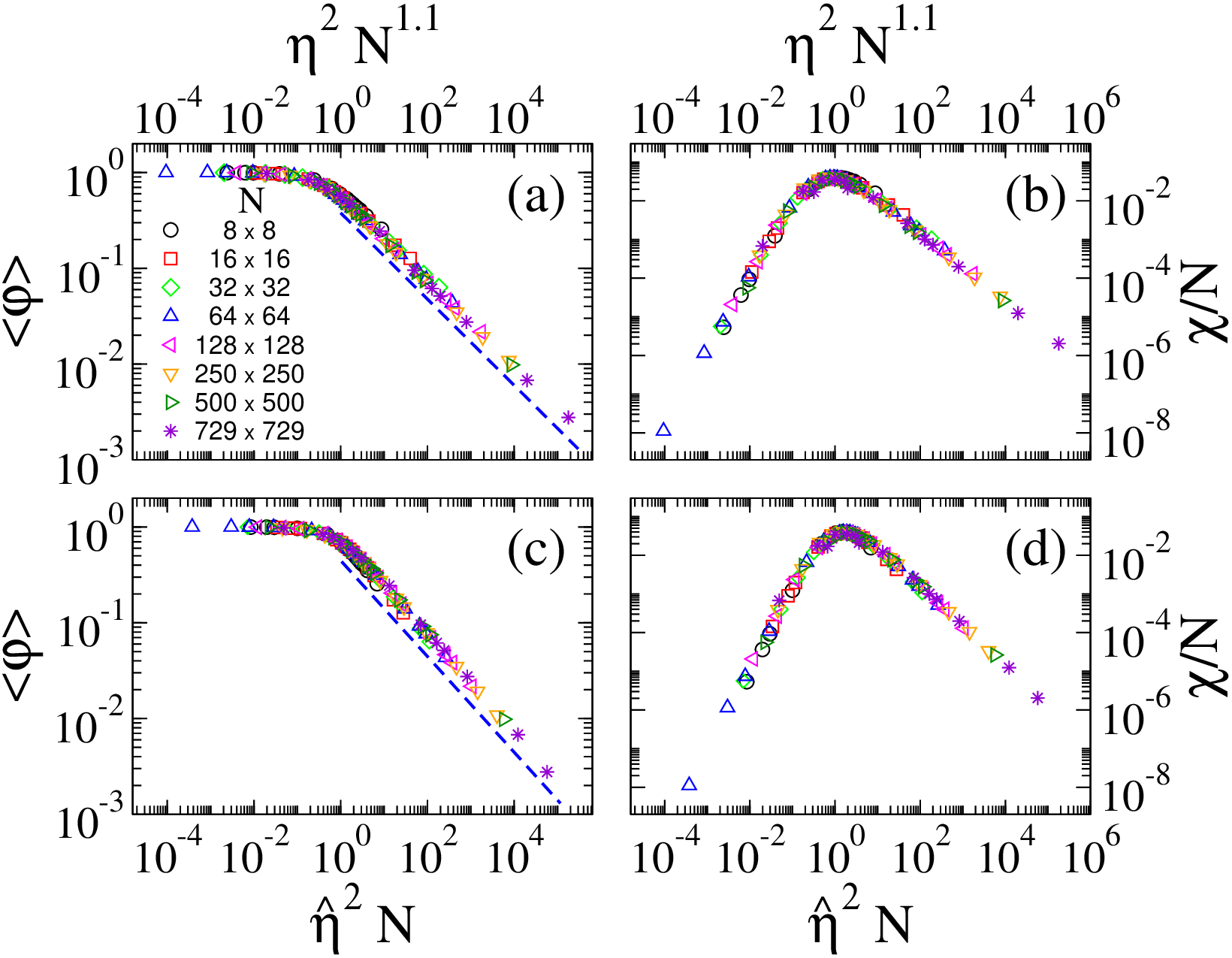}
	\caption {Static version of the FVM in two dimensional lattices.  (a) Average order parameter $\langle \varphi \rangle$ and (b) normalized susceptibility $\chi/N$ vs the scaling variable $\eta^2 N^{1.1}$ for the system sizes $N$ indicated in the legend.  (c) 
$\langle \varphi \rangle$ and (d) $\chi/N$ vs the scaling variable $x_{\mbox{\tiny 2D}} = \hat{\eta}^2 N$ for the same system sizes as in panels (a) and (b), with $\hat{\eta}=\eta \sqrt{-\ln \eta}$.  Dashed lines in panels (a) and (c) have slopes $-0.45$ and $-1/2$, respectively.}
	\label{phi-chi-2D}
\end{figure}
 
We repeated the same analysis for the FVM model on two dimensional lattices.  Simulation results are presented in Fig.~\ref{phi-chi-2D}, where the data collapse was obtained by means of two different scaling variables, as we describe below.  As it happens for the MF and the $1D$ cases, the transition noise (given by the maximum of the susceptibility) decays as a power law with the system size $N$ as $\eta_c^{\mbox{\tiny 2D}} \sim N^{-\alpha}$ [square symbols in Fig.~\ref{etac-chimax}(a)], with a best power-law fitting exponent $\alpha \simeq 0.56 \pm 0.01$.  Even though this exponent is different from the MF and $1D$ exponents $0.5$ and $1$, respectively, this numerical scaling implies an extrapolated transition noise $\eta_c^{\mbox{\tiny 2D}} = 0$ in the $N \to \infty$ limit.  The peak of the susceptibility $\chi^{\mbox{\tiny max}}$ seems to increase linearly with $N$ as in MF and $1D$, with a best-fitting exponent $\gamma \simeq 0.99 \pm 0.01$ [Fig.~\ref{etac-chimax}(b)].  Based on these results, we plot 
$\langle \varphi \rangle$ and $\chi/N$ as a function of $\eta^2 N^{1.1}$ in Figs.~\ref{phi-chi-2D}(a) and (b), respectively, where we observe a good collapse of curves for different system sizes.  For the sake of simple comparison, we have also collapsed the same data using the MF scaling variable $\eta^2 N$ instead, and found that the data points do not fall into a single curve but they look rather disperse (plot not shown).  Therefore, we conclude that the $2D$ case appears to have its own scaling variable, which is proportional to a non-trivial power of $N$.

A more appealing scaling variable can be obtained from known results of the behavior of the surface-reaction model introduced by Fichthorn, Gulari and Ziff (FGZ) in \cite{Fichthorn-1989} and studied later in  \cite{Clement-1991a,Clement-1991a,Clement-1991b,Flament-1992}, akin to the two-state NVM \cite{Kirman-1993,Granovsky-1995}, which be believe it belongs to the same class of the MSVM for flocking studied here.  In the FGZ model, $N$ particles of two different species $A$ and $B$ occupy the sites of a square lattice that simulates a catalytic substrate.  In a single step, two possible reaction events can take place.  (i) With probability $p_d$ one particle is chosen at random and desorbs, and the vacant site is immediately occupied with a particle of species $A$ or $B$ with the same probability $1/2$.  This corresponds to the external noise of the NVM that switches the state of a particle with probability $p_d/2$.  (ii) With the complementary probability $1-p_d$ a pair of neighboring sites is chosen at random and, if it is an $AB$ pair, both particles desorb and are replaced with an $AA$ or a $BB$ pair, equiprobably.  This represents the copy dynamics of the NVM.  The control parameter of the FGZ model is the desorption probability $p_d$ (noise amplitude). The steady state at $p_d=0$ is a poisoned absorbing state with a coverage equals to $1.0$ (all particles in state $A$ or $B$), which is analogous to complete order for $\eta=0$ in the FVM.  For $p_d>0$ the coverage is smaller than $1.0$, depending on the values of $p_d$ and $N$, similarly to the partial order in the FVM. 
  
It turns out that the scaling variables that we obtained for the FVM in MF and $1D$ are the same as those of the FGZ model, by making a suitable change of variables.  In the FGZ model they obtained analytically the scaling variables $X_{\mbox{\tiny MF}}=p_d \, N$ in MF ($d=3$) and $X_{\mbox{\tiny 1D}}=p_d^{1/2} N$ in $1D$ \cite{Clement-1991a,Clement-1991b}, while in the FVM are $x_{\mbox{\tiny MF}}=\eta^2 N$ in MF and $x_{\mbox{\tiny 1D}}=\eta N$ in $1D$.  Thus, the scaling variables of both models match if we make the  substitution $p_d \to \eta^2$.  Finally, $2D$ is a marginal dimension in the FGZ model, with a scaling variable similar to that of MF with a logarithmic correction in $p_d$, that is, 
$X_{\mbox{\tiny 2D}}=p_d \ln(1/p_d) N$.  Therefore, for the FVM in $2D$ we expect a scaling variable $x_{\mbox{\tiny 2D}}= \hat{\eta}^2 N$, where we have defined an effective noise amplitude $\hat{\eta} \equiv \eta \sqrt{-\ln \eta}$.   

Panels (c) and (d) of Fig.\ref{phi-chi-2D} show $\langle \varphi \rangle$ and $\chi/N$ plotted as a function of the scaling variable $x_{\mbox{\tiny 2D}}$, where we see a good data collapse.  Even though this collapse with $x_{\mbox{\tiny 2D}}$ seems as good as that with $\eta^2 N^{1.1}$ [panels (a) and (b)], the advantage of using $x_{\mbox{\tiny 2D}}=\hat{\eta}^2 N$ is two fold: we are not fitting any parameter and we recover the linear dependence on $N$ found in MF and $1D$ scaling variables $x_{\mbox{\tiny MF}}$ and $x_{\mbox{\tiny 1D}}$.  Additionally, Fig.~\ref{phi-chi-2D}(c) shows that the order parameter scales as $\langle \varphi \rangle_{\mbox{\tiny 2D}} \sim x_{\mbox{\tiny 2D}}^{-1/2}$ for $x_{\mbox{\tiny 2D}} \gtrsim 1$ (dashed line), consistent with the power law decay found in MF and $1D$.  In comparison, $\langle \varphi \rangle_{\mbox{\tiny 2D}}$ decays as a power law of $\eta^2 N^{1.1}$ with a non-trivial exponent $-0.45$ [dashed line in Fig.~\ref{phi-chi-2D}(a)].  Finally, from the scaling relation for the susceptibility
\begin{equation}
	\chi_{\mbox{\tiny 2D}} = N g_2( \hat{\eta}^2 N),
	\label{chi-eta-N-2D}
\end{equation}
where $g_2$ is a smooth function of $x_{\mbox{\tiny 2D}}$ [see Fig.~\ref{phi-chi-2D}(d)], we obtain the effective transition noise 
\begin{equation}
	\hat{\eta}_c^{\mbox{\tiny 2D}} \simeq B \, N^{-1/2}
	\label{hat-etac-N-2D}
\end{equation}
in $2D$, where $B$ is a proportionality constant.  Interestingly, the exponent $\hat{\alpha}^{\mbox {\tiny 2D}} \equiv 1/2$ in the $2D$ case agrees with that of the MF case [Eq.~(\ref{etac-N-MF})].  In the inset of Fig.~\ref{etac-chimax}(a) we compare the effective transition noise 
\begin{equation}
	\hat \eta_c^{\mbox{\tiny 2D}} = \eta_c^{\mbox{\tiny 2D}} \sqrt{-\ln \eta_c^{\mbox{\tiny 2D}}}
	\label{hat-etac-N}
\end{equation}
from simulations (circles) with the approximate scaling given by  Eq.~(\ref{hat-etac-N-2D}) (upper solid line), with a best fitting constant $B=1.3 \pm 0.04$. The good agreement between simulations and Eq.~(\ref{hat-etac-N-2D}) shows that the transformation of the original noise $\eta_c^{\mbox{\tiny 2D}}$ into the effective noise $\hat \eta_c^{\mbox{\tiny 2D}}$ leads to power-law decay in $N$ with a MF exponent $\hat{\alpha}^{\mbox {\tiny 2D}} = 1/2$. 
 
We can now obtain an approximate expression for the transition noise assuming that it has the power law behavior $\eta_c^{\mbox{\tiny 2D}} \simeq A \, N^{-\alpha}$ as found numerically [squares in Fig.~\ref{etac-chimax}(a)], where the exponent $\alpha$ depends on $N$ and $A \simeq 0.96$ is a constant obtained from the fitting of the data.  Starting from the relation Eq.~(\ref{hat-etac-N}) between the effective and original noise, we apply the logarithm at both sides and replace $\ln \hat{\eta}_c^{\mbox{\tiny 2D}}$ by $\ln B -\frac{1}{2} \ln N$ from Eq.~(\ref{hat-etac-N-2D}) and $\ln \eta_c^{\mbox{\tiny 2D}}$ by $\ln A -\alpha \ln N$, which leads to 
\begin{equation}
	(2 \alpha-1) \ln N - 2 \ln \left( A/B \right) - \ln \left( \ln N \right) - \ln \alpha = 0,
	\label{alpha-lnN}		
\end{equation}
after doing some algebra and rearranging terms.  We have also considered the expansion 
$\ln \left( -\ln A + \alpha \ln N \right) = \ln \alpha + \ln(\ln N) + \mathcal O\left[ \left( \ln A \right)/(\alpha \ln N) \right]$ to zero-th order in $\left( \ln A \right)/(\alpha \ln N) \ll 1$, as we can check for $N \gtrsim 10^2$, $A \simeq 0.96$, and $\alpha \gtrsim 1/2$.  Then, as we expect $\alpha$ to be similar to $1/2$ ($\alpha \simeq 0.56$ from the fitting of the $2D$ data in Fig.~\ref{etac-chimax}), we replace $\ln \alpha$ in Eq.~(\ref{alpha-lnN}) by the Taylor expansion $\ln \alpha \simeq \ln(1/2) + 2 \alpha -1$, and solve for $\alpha$.  We finally arrive at the following approximate scaling for the transition noise with $N$:
\begin{subequations}
  \begin{alignat}{3}
  	\eta_c^{\mbox{\tiny 2D}} &\simeq A \, N^{-\alpha}, ~~~ \mbox{with} \\
	\label{exp-alpha}
	\alpha(N) &\simeq \frac{1}{2} + \frac{ \ln \left[ \frac{A}{B} \left( \frac{1}{2} \ln N \right)^{1/2} \right]}{\ln N -1} ~~~ \mbox{or} \\ 
	\label{etac-N-lnN}	 
    	\eta_c^{\mbox{\tiny 2D}} &\simeq 1.8  \left( N \ln N \right)^{-1/2} ~~~ \mbox{for $N \gg 1$},
  \end{alignat}
  \label{etac-alpha-2D}
\end{subequations}  
using $B \simeq 1.3$.  In the inset of Fig.~\ref{etac-chimax} we can see that the approximation from Eq.~(\ref{etac-N-lnN}) (bottom solid curve) reproduces very well the behavior of $\eta_c^{\mbox{\tiny 2D}}$ vs $N$ from simulations (squares).  The second term in the exponent $\alpha(N)$ [Eq.~(\ref{exp-alpha})] leads to a very slow curvature in log-log scale with an effective exponent $\alpha \simeq 0.56$ in the shown range of $N$, which approaches very slowly to the value $1/2$ as $N$ increases.  Finally, from Eq.~(\ref{etac-N-lnN}) we can see that the transition point $\eta_c^{\mbox{\tiny 2D}}$ vanishes in the $N \to \infty$ limit.

Summarizing the results of this section, the static version of the FVM in one and two-- dimensional lattices exhibits an order-disorder transition at zero noise in the thermodynamic limit.

\section{Dynamic case $v > 0$ in two dimensions}
\label{dynamic}

When particles are allowed to move over the space, their speed $v$ becomes a relevant parameter that drastically changes the behavior of the system respect to the static case analyzed in section~\ref{static}, as we shall see below.  Simulations were done on a two--dimensional continuous space (square box) using the parallel dynamics defined in section~\ref{model}.  We remark that interactions are local, that is, each particle can only interact with other particles that are less than a distance $R=1$ apart, by copying the direction of one of them chosen at random.

\begin{figure}[t]
  \begin{center}
    \vspace{0.5cm}
    \begin{tabular}{cc}
      \hspace{-1.5cm}
      \includegraphics[width=5.3cm, bb=70 -20 550 550]{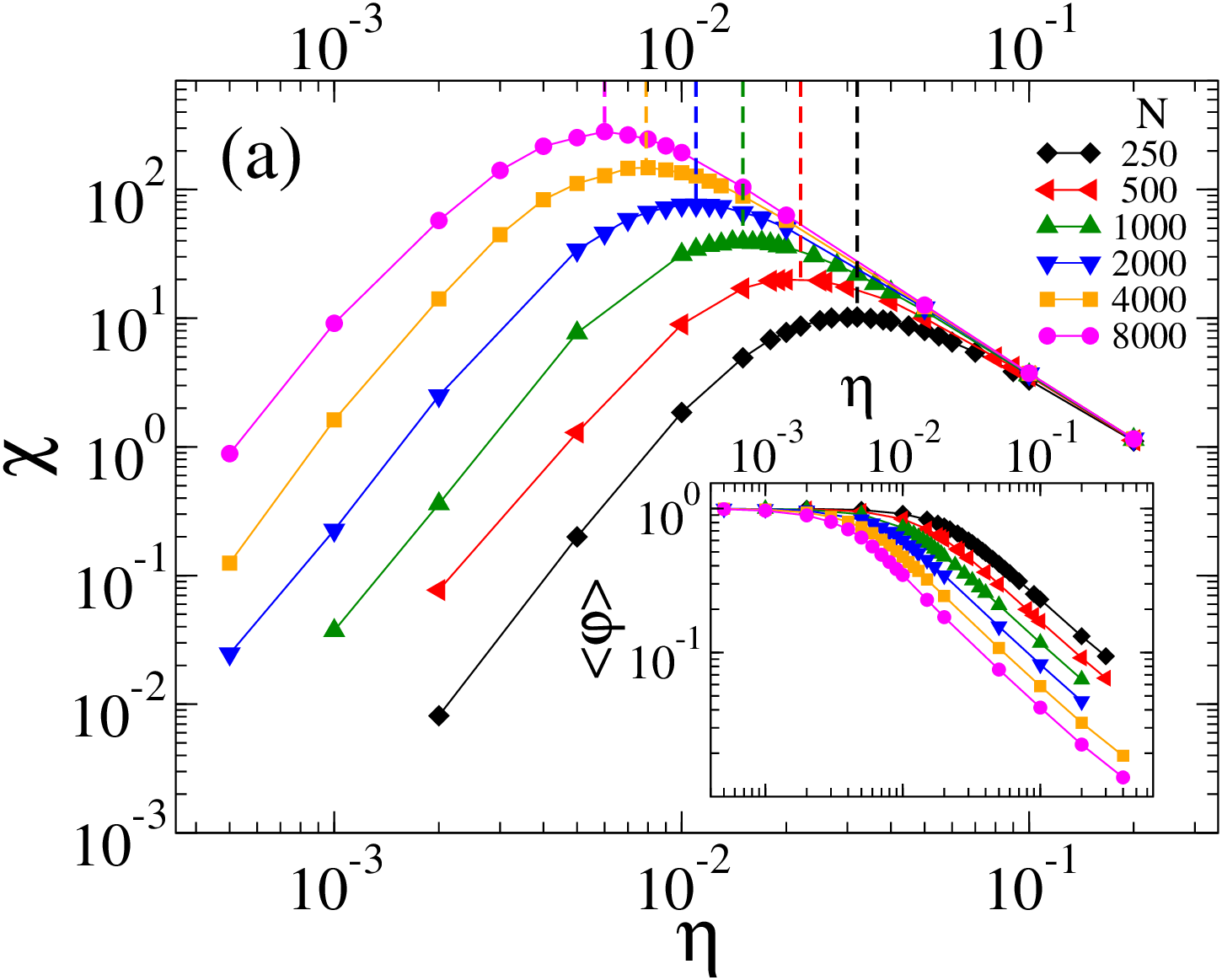} 
      & \hspace{3.0cm}      
      \includegraphics[width=5.3cm, bb=70 -20 550 550]{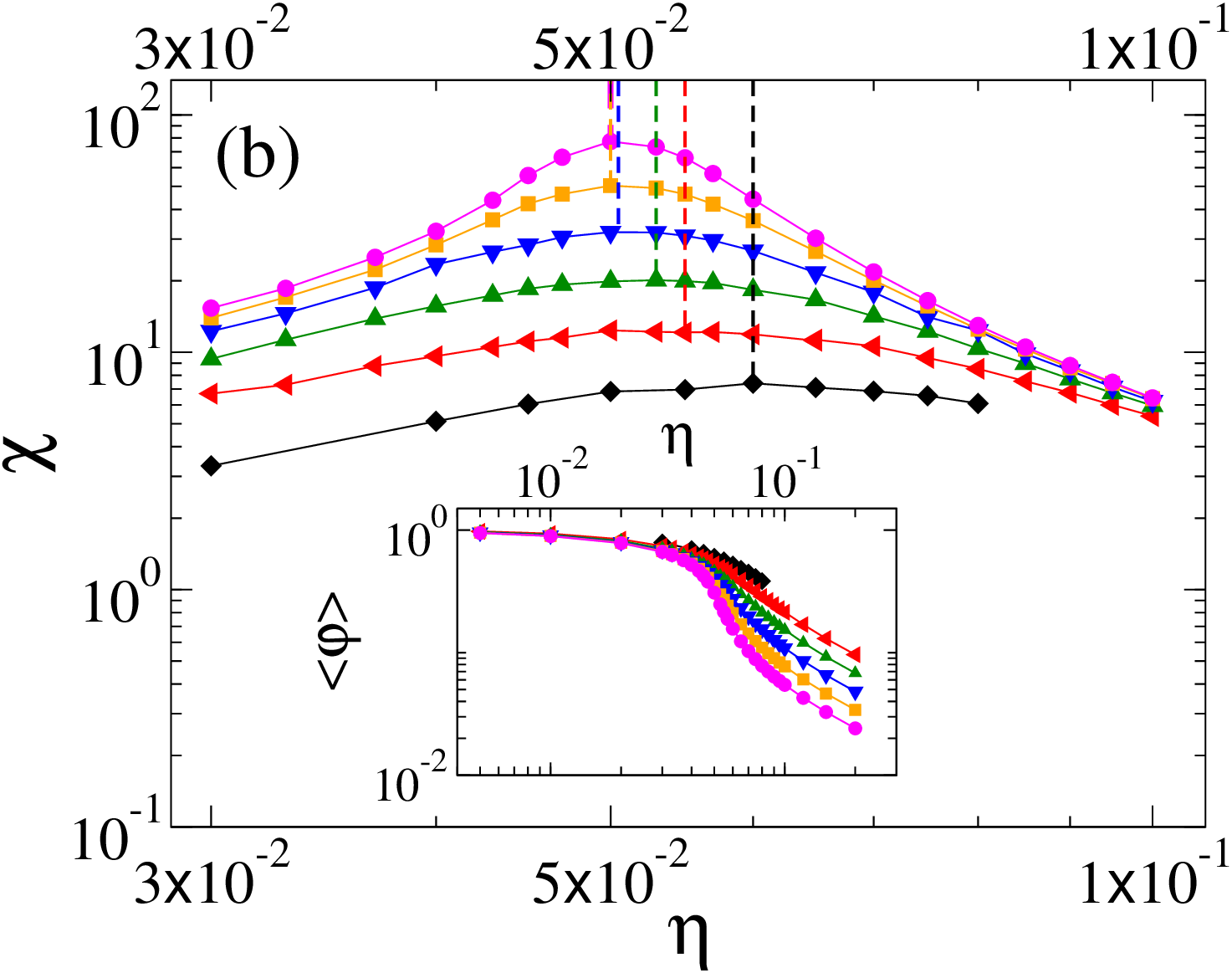} 
    \end{tabular}    
    \caption{Results of the dynamic version of the FVM in a two--dimensional continuous space with particles' speed $v=0.1$ (a) and $v=1.0$ (b).  The main panels show the susceptibility $\chi$ vs noise amplitude $\eta$ for the system sizes indicated in the legend.  The insets show the average of the order parameter $\langle \varphi \rangle$ vs $\eta$.  Vertical dashed lines indicate the estimated location of the transition noise $\eta_c$ (maximum of $\chi$).}
    \label{chi-eta-v}
  \end{center}
\end{figure}

In Fig.~\ref{chi-eta-v} we plot the susceptibility $\chi$ and the order parameter $\langle \varphi \rangle$ (inset) vs noise amplitude $\eta$ for speeds $v=0.1$ [panel (a)] and $v=1.0$ [panel (b)].  In principle, we observe a behavior similar to that of MF and the $1D$ and $2D$ static cases studied previously where $\langle \varphi \rangle$ decays monotonically with $\eta$, and $\chi$ exhibits a maximum at a value $\eta_c$ that decreases with $N$, as we can clearly see for $v=0.1$.  However, an inspection of the $v=1.0$ plot reveals that $\eta_c$ appears to decrease and saturate at a minimum value $\eta_c \simeq 0.05$ as $N$ increases, unlike in MF and the static cases where $\eta_c$ vanishes with $N$.  Also, if we compare the level of order $\langle \varphi \rangle$ and its fluctuations $\chi$ for the two speeds, we can see a larger order with smaller fluctuations for the largest speed $v = 1.0$, suggesting that the speed has an ordering effect.

\begin{figure}[t]
  \begin{center}
    \vspace{0.5cm}
    \begin{tabular}{cc}
      \hspace{-1.5cm}
      \includegraphics[width=5.3cm, bb=70 -20 550 550]{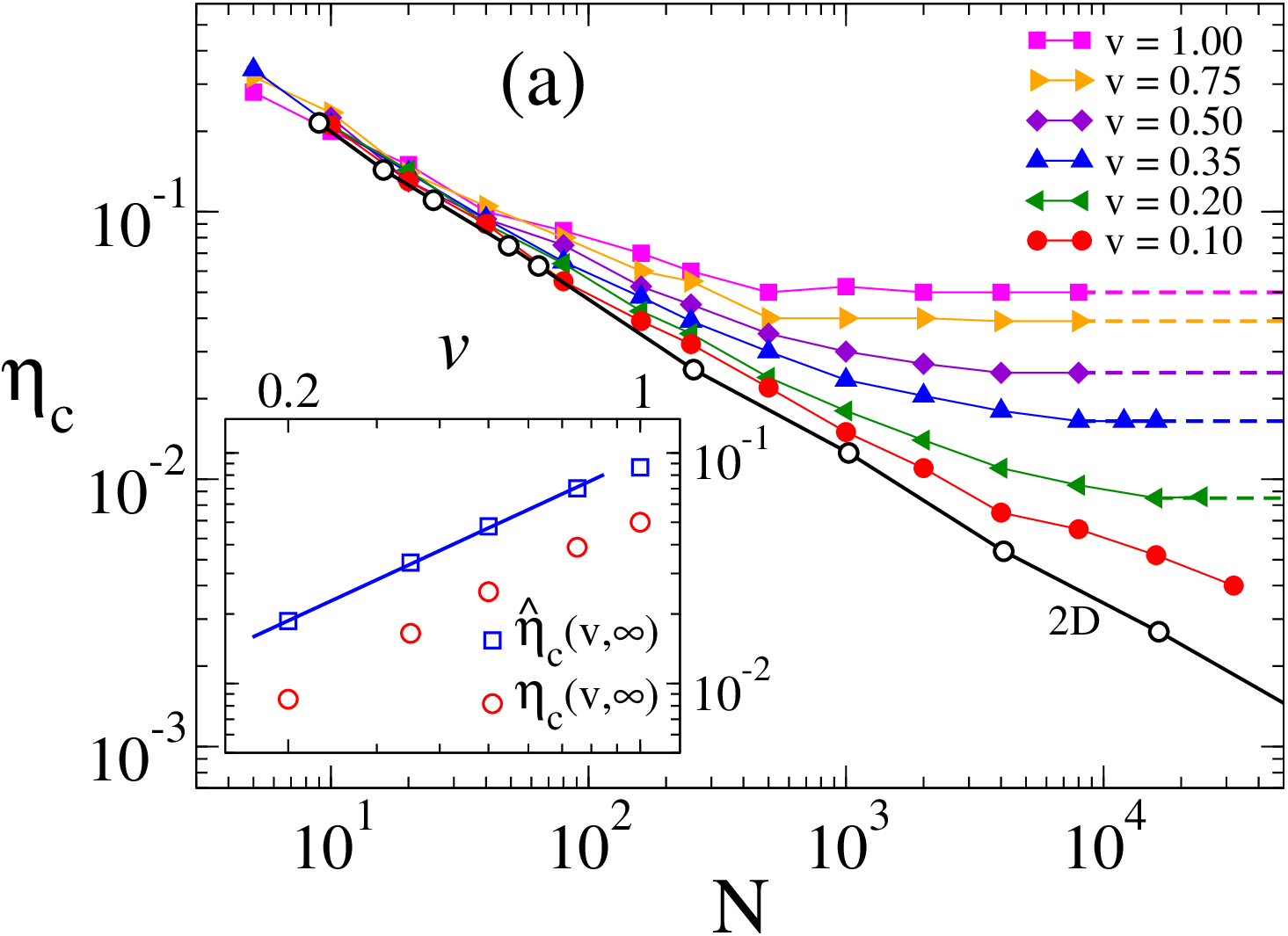}      
      & \hspace{3.0cm}      
      \includegraphics[width=5.3cm, bb=70 -20 550 550]{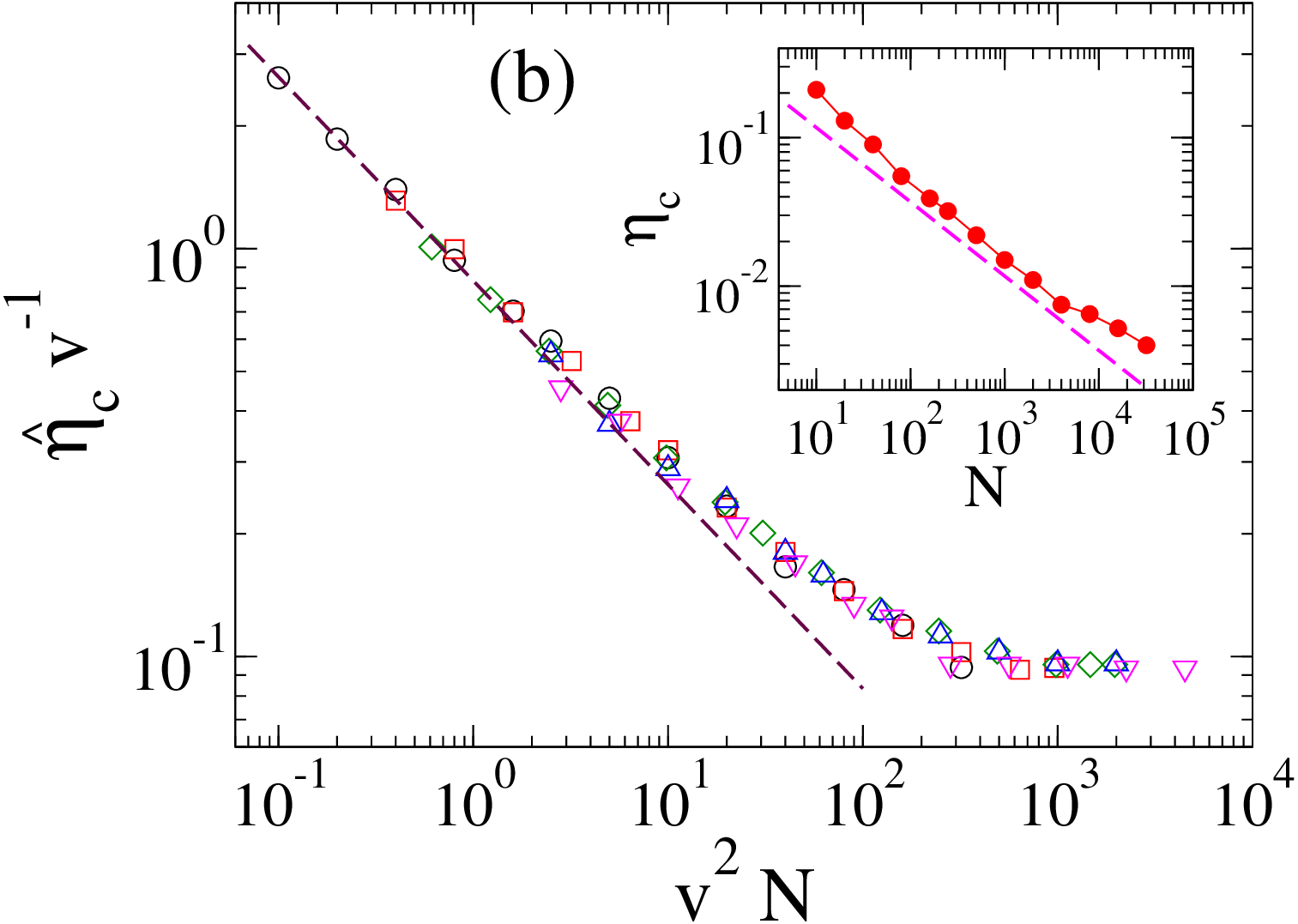} 
    \end{tabular}   
    \caption {(a) Transition noise $\eta_c$ vs system size $N$ for the speeds $v$ indicated in the legend.  The empty circles correspond to the two--dimensional static case ($v=0$) on square lattices.  The horizontal dashed lines indicate the asymptotic values $\eta_c(v,\infty)$ for large $N$.  Inset:  transition noise $\eta_c(v,\infty)$ vs $v$ (circles) obtained from the main panel, and effective transition noise $\hat{\eta}_c(v,\infty)$ vs $v$ (squares).  The straight line is best power-law fit $C \, v^{\beta}$ of $\hat{\eta_c}(v,\infty)$ for $v \le 0.75$, with resulting constant $C=0.095 \pm 0.01$ and exponent $1.01 \pm 0.02$.  (b) Collapse of the curves for the different speeds of panel (a) by means of $\hat{\eta}_c(v,N)$.  The exponents $z=2$ and $\beta=1$ in the $x$ and $y$--axis, respectively, correspond to the scaling Eq.~(\ref{hat-eta-v-N-2}). The dashed line with slope $-1/2$ indicates the power law regime for $v^2 N \lesssim 2$.  The inset shows $\eta_c$ vs $N$ for $v=0.1$.  The dashed line has slope $-1/2$.}
    \label{etac-N-v}
  \end{center}
\end{figure}

To look at this in more detail, we plot in Fig.~\ref{etac-N-v} the transition noise $\eta_c(v,N)$ vs the system size $N$ for different speeds.  Indeed, for a given speed $v \gtrsim 0.2$, we can see that $\eta_c$ exhibits a decay similar to a power law for small values of $N$, and saturates at a minimum value $\eta_c(v,\infty)>0$ for large $N$, which decreases as $v$ decreases.  We also plot for comparison the transition noise $\eta_c^{\mbox{\tiny 2D}}(N)$ for the static case $v=0$ in two--dimensional lattices (empty circles).  For the sake of clarity, the dashed line has been shifted in the $y$-axis to match the estimated asymptotic behavior of $\eta_c(v,N)$ in the zero speed limit $v \to 0$, as we do not expect $\eta_c^{\mbox{\tiny 2D}}(N)$ and $\eta_c(0,N)$ to be exactly the same.  This is because some macroscopic magnitudes of the dynamic model ($\langle \varphi \rangle$, $\chi$ and $\eta_c$) depend on other variables besides $v$ and $N$, such as the density of particles $\rho$.

The numerical results described above show that, in the thermodynamic limit, there is an order-disorder transition at a finite noise amplitude $\eta_c>0$ that increases with the speed $v$.  To study this transition in more detail, we investigate below the scaling behavior of $\eta_c$ with the speed and the system size.

Since we have learned in section~\ref{static} that working with an effective noise $\hat{\eta}$ in $2D$ lattices leads to scalings with simple MF exponents, it seems reasonable to explore the data of Fig.~\ref{etac-N-v} for an effective transition noise 
\begin{equation}
	\hat{\eta}_c(v,N) \equiv \eta_c(v,N) \sqrt{-\ln \eta_c(v,N)},
	\label{hat-etac-etac} 
\end{equation}
which incorporates a correction factor $\sqrt{-\ln \eta_c}$ to the original noise $\eta_c$.  The approximate power-law decay of $\eta_c$ for small $N$ and its saturation for large $N$ [Fig.~\ref{etac-N-v}(a)] suggests that the scaling behavior of $\hat{\eta}_c(v,N)$ could be described by the following standard Family-Vicsek function with two independent exponents $\beta$ and $z$ \cite{Family-Vicsek-1985}:
\begin{equation}
	\hat{\eta}_c(v,N) \sim v^{\beta} f \left( v^z N \right),
\label{eta-v-N}
 \end{equation}
 where $f$ is a scaling function with the asymptotic properties
\begin{eqnarray}
  f(x) \sim 
  \begin{cases}
    x^{-\alpha} & \mbox{for $x \ll 1$}, \\
    \mbox{constant} & \mbox{for $x \gg 1$}.
  \end{cases}
  \label{f}
\end{eqnarray}
We can check that Eq.~{(\ref{eta-v-N})} exhibits the two limiting behaviors
\begin{equation}
	\hat{\eta}_c(v,N \to \infty) \sim v^{\beta} 
\label{Eq:beta2}
\end{equation}
in the thermodynamic limit, and 
\begin{equation}
	\hat{\eta}_c(v \to 0, N) \sim N^{-\alpha}
\label{nu2}
 \end{equation}
 in the zero speed limit, where the exponent $\alpha$ satisfies the relation 
\begin{equation}
	\beta=z \, \alpha.
\label{beta-z-alpha}
\end{equation}
By means of the scaling relation Eq.~(\ref{eta-v-N}) we can collapse the data points of Fig.~\ref{etac-N-v} into a single curve.  For that, we first estimate the exponents $\beta$, $\alpha$ and $z$.  From the plot $\hat{\eta}_c(v,\infty)$ vs $v$ in the inset of Fig.~\ref{etac-N-v} (squares) we find the best power-law fitting $C \, v^{\beta}$ (straight line), where $C=0.095 \pm 0.01$ and $\beta=1.01 \pm 0.02$.  Then, in the zero speed limit we assume that $\alpha$ takes the value $\alpha=\alpha^{\mbox{\tiny 2D}}=1/2$ of the $2D$ static case, and thus we obtain $z = 2.02 \pm 0.04$ from Eq.~(\ref{beta-z-alpha}).  Based on these exponents, we propose the following scaling for the effective transition noise:
\begin{equation}  
	\hat{\eta}_c(v,N) \sim v \, f \left( v^2 N \right),
	\label{hat-eta-v-N-2}	
\end{equation}
with $f(x) \sim x^{-1/2}$ for $x \ll 1$ and $f(x) \sim \mbox{const}$ for $x \gg 1$.  Figure~\ref{etac-N-v}(b) shows a good data collapse obtained with the scaling Eq.~(\ref{hat-eta-v-N-2}).  Remarkably, this result only required the estimation of the best fitting exponent $\beta$ of the $\hat{\eta}_c(v,\infty)$ vs $v$ data, and assuming that the scaling of the transition noise with N in the zero speed limit is the same as that of the $2D$ static case.

The effective transition noise given by Eq.~(\ref{hat-eta-v-N-2}) scales linearly with the speed in the thermodynamic limit, 
\begin{equation}
	\hat{\eta}_c(v,\infty) \simeq C \, v,
	\label{hat-etac-v} 
\end{equation}
where $C = 0.095$ is the best fitting constant for low speeds $v \lesssim 0.75$ [straight line in the inset of Fig.~\ref{etac-N-v}(a)].  An approximate power-law scaling $\eta_c(v,\infty) \simeq D \, v^{\beta}$ for the original noise  can be obtained by following the same approach described in section~\ref{static} to obtain the scaling of $\eta_c^{\mbox{\tiny 2D}}$ with $N$ [Eq.~(\ref{etac-alpha-2D})].  For that, we start from the relation between $\hat{\eta}_c$ and $\eta_c$ in logarithmic scale $\ln \hat{\eta}_c = \ln \eta_c + (1/2) \ln( -\ln \eta_c)$ and replace $\ln \hat{\eta}_c$ by $\ln C + \ln v$ [Eq.~(\ref{hat-etac-v})] and $\ln \eta_c$ by $\ln D + \beta \ln v$.  After rearranging terms and making the approximation $\ln \left( -\ln D - \beta \ln v \right) \simeq \ln \beta + \ln(-\ln v)$ to zero-th order in $(\ln D) / (\beta \ln v) < 1$ we arrive at
\begin{equation}
	2(\beta-1) \ln v - 2 \ln(C/D) + \ln(-\ln v) + \ln \beta = 0. 
	\label{eqn-v}
\end{equation}
As we expect $\beta$ to be similar to $1.0$ [circles in the inset of Fig.~\ref{etac-N-v}(a)], we use the linear approximation $\ln \beta \simeq \beta -1$ in Eq.~(\ref{eqn-v}) and solve for $\beta$.  We finally obtain the following approximate expressions for the transition noise: 
\begin{subequations}
  \begin{alignat}{2}
  	\eta_c(v,\infty) &\simeq D \, v^{\beta}, ~~~ \mbox{with} \\
	\label{exp-beta}
	\beta(v) &\simeq 1 + \frac{\ln \left[ \frac{C}{D} \left( -\ln v \right)^{-1/2} \right]}{\ln v +1/2} ~~~ \mbox{or} \\
	\label{etac-v-app}
	\eta_c(v,\infty) &\simeq C \, v \left( -\ln v \right)^{-1/2} ~~~ \mbox{for $v \ll 1$}.
  \end{alignat}
  \label{etac-beta}
\end{subequations}  
The second term in Eq.~(\ref{exp-beta}) gives an effective exponent $\beta(v) \gtrsim 1$ that decreases and approaches the value $1$ very slowly as $v$ decreases. Equations~(\ref{etac-beta}) are only valid for low speeds due to the fact that the approximate expansion of the logarithm that we used in Eq.~(\ref{eqn-v}) assumes that $(\ln D) / (\beta \ln v) < 1$, which happens for $v \lesssim 0.08$.  Unfortunately, the comparison of Eq.~(\ref{etac-beta}) with simulation results is not possible because to obtain the numerical value $\eta_c(v,\infty)$ for speeds $v < 0.2$ is extremely costly in terms of simulation running times. 

Equation~(\ref{hat-eta-v-N-2}) also implies the scaling 
\begin{equation}
	\hat{\eta}_c(v,N) \sim N^{-1/2} ~~~ \mbox{for $v^2 N \ll 1$},
\label{nu2}
 \end{equation}
which is confirmed in Fig.~\ref{etac-N-v}(b), where the collapsed data exhibits an approximate power law decay with exponent $-1/2$ for $v^2 N \lesssim 2$, denoted by the dashed line.  Finally, in the inset of Fig.~\ref{etac-N-v}(b) we compare the curve $\eta_c$ vs $N$ for the lowest speed $v=0.1$ with the $N^{-1/2}$ scaling (dashed line).  A good agreement is observed only at intermediate values of $N$, while for small or large sizes a deviation from the slope $-1/2$ becomes clear.  We understand that the discrepancy for small $N$ is due to the absence of the logarithmic correction $\sqrt{-\ln \eta_c}$ that becomes more relevant as $\eta_c$ decreases, while for large $N$ we expect that $\eta_c$ reaches a saturation at a minimum value $\eta_c(0.1,\infty)>0$.  This asymptotic value of $\eta_c(0.1,N)$ is reached for system sizes outside the shown range and, in general, the approximate system size from where we start to see a plateau in $\eta_c$ seems to diverge as $v$ approaches zero [see Fig.~\ref{etac-N-v}(a)].  An insight into this can be given in terms of the crossover size $N_{\mbox{\tiny cross}}$ that separates the two limiting behaviors of $\eta_c(v,N)$ for small and large $N$.  For $N \ll N_{\mbox{\tiny cross}}$ the effective transition noise decays with $N$ as $\hat{\eta}_c \sim N^{-1/2}$, while for $N \gg N_{\mbox{\tiny cross}}$ is $\hat{\eta}_c \sim v$.  At the crossover size, these two limiting scalings should match, leading to $N_{\mbox{\tiny cross}} \sim v^{-2}$.  This  simple relation shows that, as $v$ approaches zero, the crossover size diverges very fast, and so we need to run simulations in very large systems to observe the asymptotic value of $\eta_c(v,N)$.
 
In summary, we showed in this section that the FVM in a $2D$ continuous space exhibits and order-disorder phase transition at a finite noise amplitude $\eta_c>0$  that is proportional to the speed $v$ of particles.  For low speeds, $\eta_c$ is linear in $v$ with a logarithmic correction that leads to an effective power law with a $v$--dependent exponent slightly larger than $1$.  Thus, the transition at a finite noise $\eta_c>0$ induced by particles' motion is in contrast with the zero-noise transition found in MF and the static version of the model in lattices.

\section{Summary and Conclusions}
\label{conclusions}

We studied a model for the flocking dynamics of self-propelled particles with pairwise copying interactions and noise.  This model can be considered as a version of the noisy voter model with infinite number of angular states, which also incorporates the motion of particles over the space.  We focused on the ordering properties of the system by exploring the order parameter $\varphi$ that measures the global level of alignment of particles.  We found that the system undergoes a transition as the noise amplitude $\eta$ overcomes a threshold $\eta_c$, from an ordered phase for $\eta < \eta_c$ where a fraction of particles are aligned and thus $\varphi >0$, to a disordered phase for $\eta > \eta_c$ characterized by each particle moving in a  random direction, leading to $\varphi=0$.  We performed a numerical analysis to investigate how the speed of particles, the space and its dimension affect the order-disorder phase transition.  We started by the simplest case of all-to-all interactions or infinite dimension or MF, followed by the static case of fixed particles on one and two--dimensional square lattices, and ending with the dynamic case of particles moving on a bounded continuous two--dimensional space.  The transition point $\eta_c$ was determined by the location of the peak of the susceptibility, which depends on the system size $N$.  By doing suitable finite size scaling analysis we were able to infer the scaling behavior of the relevant magnitudes in the thermodynamic limit, including the transition noise.  

In the MF case we showed that the transition noise vanishes with $N$ as $\eta_c^{\mbox{\tiny MF}} \sim N^{-1/2}$, which is related to known analytical MF results of the MSVM.  In the static case ($v=0$) we found the scalings $\eta_c^{\mbox{\tiny 1D}} \sim N^{-1}$ in $1D$ and $\hat{\eta}_c^{\mbox{\tiny 2D}} \sim N^{-1/2}$ in $2D$, where $\hat{\eta}_c^{\mbox{\tiny 2D}} = \eta_c^{\mbox{\tiny 2D}} \sqrt{-\ln \eta_c^{\mbox{\tiny 2D}}}$ is an effective noise amplitude.  This effective noise with a logarithmic correction in $\eta_c^{\mbox{\tiny 2D}}$ was found by drawing an analogy between our FVM and the FGZ model for catalytic reactions with desorption probability $p_d$, and making the transformation $p_d \to \eta^2$.  Our scaling results on MF and lattices are compatible with those predicted theoretically for the FGZ model, which is a version of the noisy two-state voter model. 

We therefore conclude that, in MF and $1D$ and $2D$ static cases, the FVM displays an order-disorder transition at zero noise in the thermodynamic limit.  This result means that any finite noise suppresses completely any level of order in the thermodynamic limit.  That is, even a tiny amount of noise is enough to bring the system to complete disorder.

The behavior of the model in the dynamic case, where particles move at a finite speed $v>0$ on a $2D$ box, is very different to that of the MF and static cases.  We observed that, for a fixed density of particles $\rho=0.5$ and a given noise $\eta>0$, increasing the speed leads to a larger value of $\varphi$ with smaller fluctuations (smaller susceptibility $\chi$), eventually inducing a stationary state of collective order for high enough speeds.  We understand that this ordering effect produced by particles' motion is analogous to that found in Vicsek type models and, as a consequence, the system exhibits an ordered phase below a finite transition noise amplitude $\eta_c(v)>0$ that depends on the speed.  For low speeds, the behavior of the effective transition noise $\hat{\eta_c}=\eta_c \sqrt{-\ln \eta_c}$ with $v$ and $N$ is well described by a scaling function with two simple exponents.  On the one hand, this leads to the scaling behavior $\hat{\eta_c} \sim N^{-1/2}$ for $v^2 N \ll 1$, which agrees with that of the $2D$ static case, and also with the theory developed for the saturation transition in the FGZ model \cite{Clement-1991a,Clement-1991b}.  On the other hand, the effective noise reaches an asymptotic value as $N$ increases, which behaves as $\hat{\eta_c} \sim v$ in the $N \to \infty$ limit.  This results in a transition noise with a superlinear dependence on the speed of the form $\eta_c \sim v (-\ln v)^{-1/2}$ for $v \ll 1$, in the thermodynamic limit.  For the sake of comparison, it was recently found that in the Vicsek model the transition noise scales as $\eta_c \sim v^{0.45}$ in the low density and low speed regime \cite{Leticia-2019}.  We also note that the transition noise for a given speed and density $\rho=0.5$ in the FVM is much smaller than that of the Vicsek model.    

In summary, we found that the collective motion of self-propelled particles on a $2D$ space with noisy voter interactions exhibits an order-disorder transition at a finite noise amplitude $\eta_c$ proportional to the speed of particles.  This is a surprising result within the literature of the voter model, as it is known that adding an external noise to the copying dynamics of the model wipes up collective order in the thermodynamic limit, and in this article we showed that order can indeed be sustained by particles' motion.  

It seems that the effect of motion is to correlate distant particles generating a state of global order, as it happens in the Vicsek model.  Thus, it might be interesting to study the correlations between particles' velocities and positions in order to understand the mechanisms that lead to flocking in the model.  We also note that the MF approximation, which predicts a transition at zero noise, fails for the full version of the FVM with particles moving at a finite speed, showing the importance of taking into account the space and motion of particles in real life situations, as it happens for instance in the recent experiments with fish \cite{jhawar2020} described in section~\ref{introduction}.  It would be worthwhile to develop a mathematical description of the FVM that goes beyond MF and accounts for correlations between particles, which could correctly capture the ordering effect of motion.  Finally, within the context of the experiments in \cite{jhawar2020}, the results we obtained in the present article suggests that a group of fish could eventually reach an asymptotic polarized state when the group size increases, depending on the relation between the amplitude of the spontaneous directional change (noise) of fish and their speed.

\section*{ACKNOWLEDGMENTS}

We acknowledge financial support from CONICET (PIP 11220150100039CO) and (PIP 0443/2014). We also acknowledge support from Agencia Nacional de Promoci\'on Cient\'ifica y Tecnol\'ogica (PICT-2015-3628) and (PICT 2016 Nro 201-0215).

\end{document}